# Green-Graphene protective overlayer on optical microfibers: prolongs the device lifetime

*Anastasia Novikova   Aviad Katiyi   Aviran Halstuch   Alina Karabchevsky**

A. Novikova, A. Katiyi, Dr. A. Halstuch, Prof. A. Karabchevsky
School of Electrical and Computer Engineering
Ben-Gurion University of the Negev
Beer-Sheva 8410501, Israel
E-mail: alinak@bgu.ac.il

Optical microfibers finds new applications in various fields of industry, which in turn requires wear resistance, environmental friendliness and ease of use. However, optical microfibers are fragile. Here we report a new method to prolong the microfiber lifetime by modifying its surface with green-extracted graphene overlayers. Graphene films were obtained by dispergation of shungite mineral samples in an aqueous medium. For this, we tapered optical fibers and sculptured them with graphene films mixed with gold nanoparticles. We observed that due to the surface modification the lifetime and survivability of the microfiber increased 5 times, as compared to the bare microfiber. The embedded gold nanoparticles can also be utilized for enhanced sensitivity and other applications.

## 1  Introduction

Optical fibers are dated back to 1880 when William Wheeler transmitted light through a glass pipe and referred to it as a 'light piping'. It took until 1966 when the optical fibers were proposed to have a higher guiding medium as compared to the cladding for light transmission [1]. This development was intended mainly for the telecommunication due to low losses and an ease in fabrication [2]. Optical fibers can be tapered to to microfiber dimensions for experiencing the novel properties [3]. Microfiber [4] can be used for a verity of applications such as sensing [5, 6, 7, 8], determination of substances [9], human health monitoring [10] and many others. One of the applications in which microfibers are important is sensing. Due to the squeezing of the fiber diameter, the confinement of the mode decreases and the evanescent field penetration depth to the analyte enhances [11, 12]. This can be utilized for sensing with tapered fibers [13, 11, 14].

These days, optical microfibers are widely used for the determination of various substances with low concentrations. Namely, in biomedicine [15] and biology [16] (low concentrations of viruses [17, 18], bacteria [19, 20], proteins [21], nucleic acids [22], cancer cells [23], substances in body [24]), environmental protection [25] (pollutants in water and soil, components of biological pollution) [26, 27, 28, 29, 30], pharmacology and the chemical industry (pharmaceutical substances, new materials) [31, 32], construction industry (concrete deformation/stress measurements [33], the use of fiber-optic sensors for detecting railway vehicles and monitoring the dynamic characteristics of the rock mass caused by railway rolling stock for the needs of civil construction [34, 35], damage detection and characterization using long-gauge and distributed fiber optic sensors [36], a new tool for temperature measurements in boreholes [37, 38]), also optical sensors are used in the creation of smart fabrics [39], stratigraphy [40, 41] (probe for rapid snow grain size, determination of layered, sedimentary and volcanogenic rocks).

The use of microfibers is effective, due to its high sensitivity to the materials under study, it is environmentally safe since microfibers consists of SiO and can be easily disposed. In addition to the above mentioned advantages, when using optical microfibers, little time is needed to determine the substances under study. However, there is a challenge related to the thickness of the microfiber. First, researchers need to be very careful when working with such a fragile device. Second, when working with metal nanoparticles and highly toxic substances, the microfiber can also be easily destroyed during the research process. When using this type of microfiber on a large scale, this disadvantages can be significant. One solution to this problem is the chemical modification of the microfiber surface with various substances [42].

Here, we report the use of graphene films [43] with gold nanoparticles to increase the lifetime of microfiber as illustrated in Fig. 1. This type of microfiber is environmentally friendly, safe, easy to use and affordable. We deposited graphene films on the surface of microfiber, increasing its lifetime, which prevents the rapid destruction of microfiber by gold nanoparticles. In addition, the graphene films do not interfere with the determination of the peaks of gold nanoparticles. An increase in the lifetime of microfiber makes it possible to use it on an industrial scale for the determination of chemicals and microbiological objects. furthermore, the embedded gold nanoparticles on the graphene can be used for enhancing the sensitivity of the microfiber.

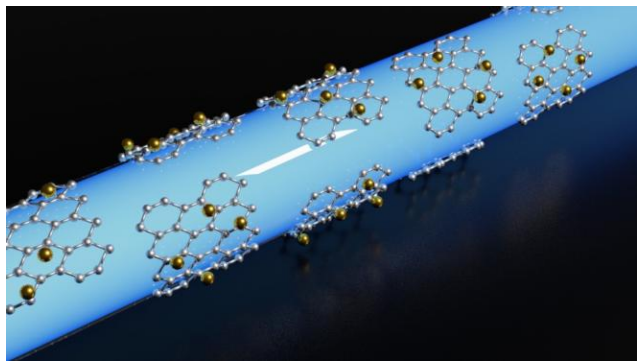

Figure 1: Illustration of a microfiber with a protective overlayer of a composite of graphene and gold nanoparticles.

## 2 Results and Discussion

To increase the lifetime of the microfiber, we used graphene films, which prevent the microfiber from contacting with gold, but do not change the peaks of gold during research. We obtained graphene films from the natural mineral shungite by the method of dispergation in an aqueous solution without the addition of surfactants. We studied the surface of shungite samples before and after dispergation treatment to determine how the surfaces of the samples changed by scanning and transmission electron microscopy.

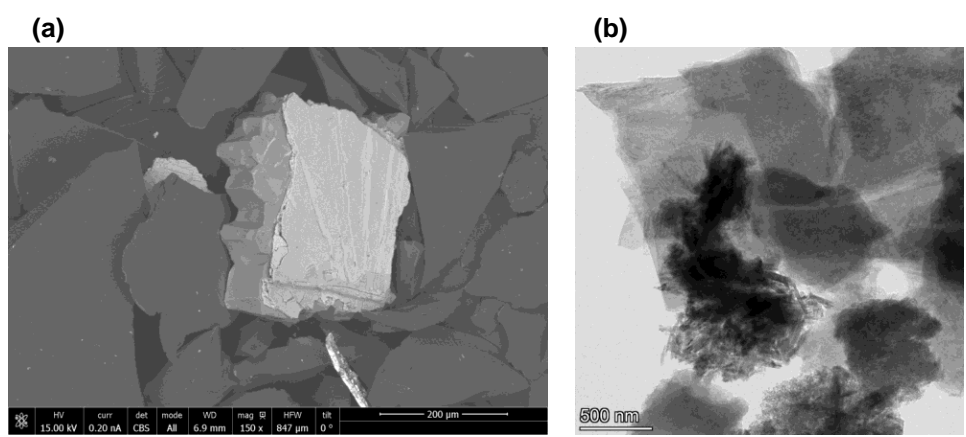

Figure 2: (a) Scanning electron microscopy (SEM) image of shungite particles before dispergation treatment and (b) transmission electron microscopy (TEM) image of graphene layers obtained from shungite.

Figure 2 shows scanning and transmission electron microscopy images of shungite particles before and after the dispergation. Figure 2a shows a scanning electron microscopy image of the shungite sample (98% of carbon in amorphous form) before despergation. It shows that shungite is homogeneous with small mineral inclusions. Figure 2b shows a transmission electron microscopy image of that shungite samples were stratified into thin graphene films, and the specific surface area of the particles increased. The sample contains graphene films and graphite-shaped parts.
For the chemical modification of the fiber surface, we use graphene with gold nanoparticles coating. We added gold nanoparticles with a diameter of 30 nm to the graphene samples obtained during dispergation. To investigate the interaction between the graphene films to the gold nanoparticles, we studied samples of graphene with gold nanoparticles and investigate it using transmission electron microscopy (TEM).

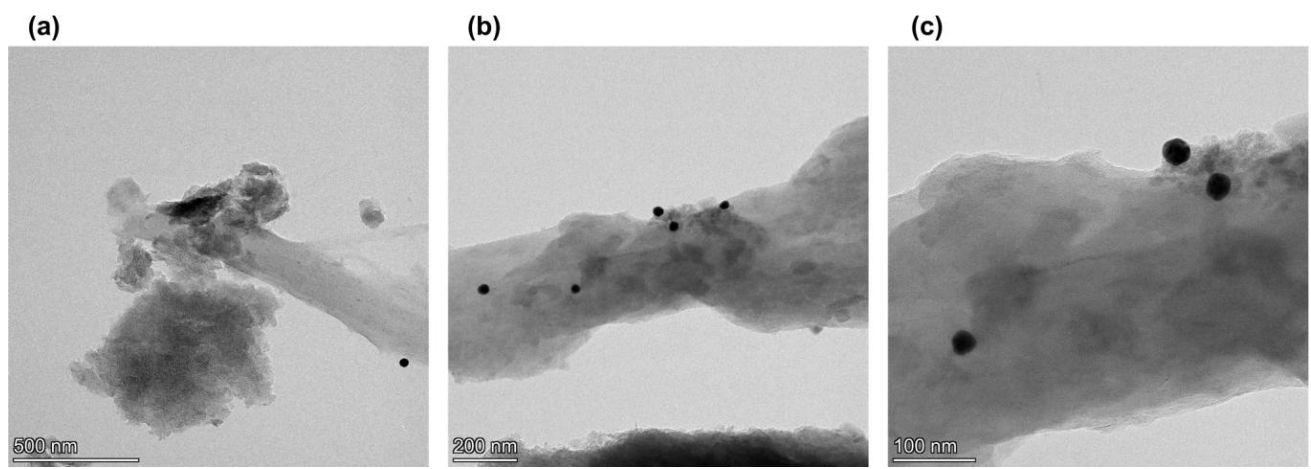

Figure 3: Transmission electron microscopy images of (a) graphene from shungite particles with gold nanoparticles with gold to shungite ratio 1:1, (b) graphene form shungite particles with gold nanoparticles with 3:1 ratio of gold to shungite and (c) zoomed image of graphene form shungite particles with gold nanoparticles with ratio of gold to shungite 3:1.

Figure 3a shows a transmission electron microscopy image of gold nanoparticles deposited on the graphene surface with a concentration of 1:1, it shows that this concentration is too low for developed graphene surfaces, we experimentally determined the optimal ratio of graphene and gold nanoparticles. Figure 3b shows TEM image when the ratio of graphene films and gold nanoparticles in the sample is increased to 1:3, it shows that gold nanoparticles are deposited on the surface. Zoom image 3c shows that the gold nanoparticles are embedded on graphene films and not separated from the graphene films. Therefore, we concluded that ratio of 1:3 is good for surface modification.

Next, we study the samples of graphene with gold nanoparticles via UV-VIS spectrometry and Raman spectroscopy to determine the spectra of graphene and gold and how they affect each other in a composite material.

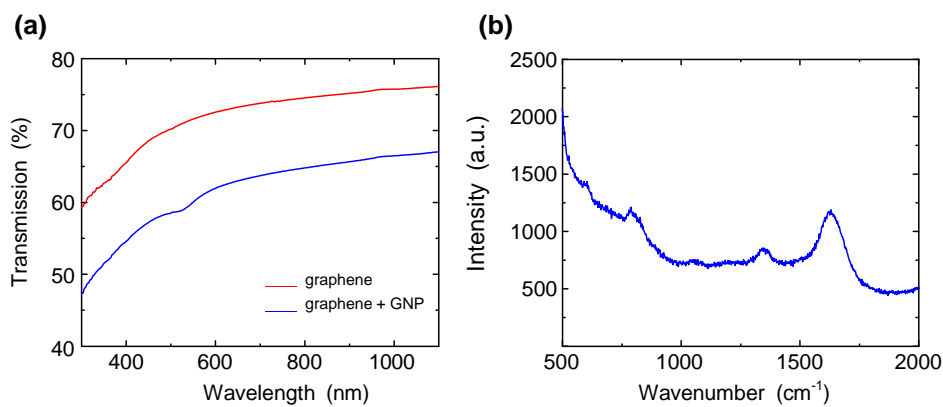

Figure 4: (a) UV-VIS spectra of graphene (red curve) and graphene with gold nanoparticles (blue curve) and (b) raman spectra of graphene with gold nanoparticles.

Figure 4a shows the transmission UV-VIS spectra of graphene and graphene with gold nanoparticles with a diameter of 30 nm. The absorption dip of the gold nanoparticles appears at 518 nm. The dip of graphene tends towards 217 nm. Figure 4b shows Raman spectra of sampled excited at 532 nm while graphene with gold nanoparticles are of 30 nm in diameter. $D1 = 1345.56$ cm$^{-1}$ (graphite form , sp3), $G1 = 1638.36$ cm$^{-1}$ (graphene form, sp2) peaks refer to graphene films doped with graphite form. The intensity of the graphene form is higher as compared to the graphite form; accordingly, the presence of graphite is insignificant. The peak of gold nanoparticles is located at 1025.94 cm$^{-1}$ while the other of the peaks are related to the water and the cuvette in which the sample was studied.

Next, we studied the samples using X-ray photoelectron spectrometry to determine the composition of samples. Data are presented in Tables 1 and 2.

Table 1: Elements composition and quantitative evaluation of samples of treated shungite

| Name | Peak (BE) | FWHM (eV) | Area (CPS·eV) | Atomic (%) |
| --- | --- | --- | --- | --- |
| C1S | 284.74 | 1.62 | 16460.49 | 56.92 |
| O1S | 532.11 | 2.33 | 32044.85 | 38.95 |
| Ti2P3 | 458.70 | 1.16 | 6125.11 | 4.13 |

Table 2: Elements composition and quantitative evaluation of samples of treated shungite with gold nanoparticles

| Name | Peak (BE) | FWHM (eV) | Area (CPS·eV) | Atomic (%) |
| --- | --- | --- | --- | --- |
| C1S | 284.45 | 1.46 | 16150.69 | 54.34 |
| O1S | 531.70 | 3.28 | 34194.57 | 40.43 |
| Au4F | 89.22 | 6.78 | 630.82 | 0.12 |
| Ti2P3 | 458.40 | 1.24 | 7768.65 | 5.10 |

Comparing the data given in Tables 1 and 2, we see that the gold nanoparticles peak appeared at 89.22 eV, Area 630.82 CPS·eV. The carbon peak at 284.74 eV and 284.45 eV refer to the graphene peak. The samples also contain small admixtures of titanium and oxygen, since the dispergation process took place in water. The full width at half maximum of the spectral line (FWHM) was 1.62 eV after the addition of gold nanoparticles changed to 1.46 eV. The percentage of gold in the total composition was 0.12 percent.

To investigate different samples with different ratios of gold nanoparticles to graphene, we built the experimental setup shown in Fig. 5a. A broadband laser was coupled to a single mode fiber via x10 objective and was aligned to maximal power. The fiber was spliced with the tapered fiber shown in Fig. 5b. The fiber was spliced with the tapered fiber sensing device as a commercial single mode fiber to a microfiber with diameter of 2.5 $\mu m$. The microfiber region acts as the sensing region for samples characterization due to the high evanescent field in this region. The output of the tapered fiber was spliced to a pigtail single mode fiber that was connected to an optical spectrum analyzer.

For verifying the optimal concentration of gold nanoparticles, we measured graphene with gold nanoparticles with different concentrations using cuvette. We measured samples with ratio of graphene to gold nanoparticles of 1:1, 1:2, 1:3 and 1:5. Figure 5c shows VIS-NIR measurements of graphene and graphene with gold nanoparticles with different ratio of 1:1, 1:2, 1:3, and 1:5. While focusing on the wavelength range of 1300-1700 nm (Figure 5d), we can see the effect of different concentrations of gold nanoparticles on the graphene peak. The peak of graphene is shown at 1680 nm. At a ratio of 1:1 graphene to gold, the intensity decreased and the peak is shifted to the right, which indicates that the concentration of gold nanoparticles is too low. At a ratio of 1:2 and 1:3 with an increase in the concentration of gold, the intensity also increased and the peaks are also shifted to the right. Therefore, the optimal ratio to increase the intensity of peaks is 1:3. Figure 5e shows two graphs measurements of graphene with gold with concentration of 1:3 graphene to gold. The absorption peak at 560 nm is associated with the surface plasmon excitation of gold nanoparticles with a diameter of 30 nm. The peak is clearly visible as in the first and second sample. This means that with this ratio of graphene to gold, graphene does not interfere with the measurement of gold peaks. After finding the optimal concentration of gold nanoparticles in graphene, we measured the transmission of the sample using a tapered fiber. A volume of 6 um was dripped on a Teflon spacer that gently slid to the microfiber region.

Since it is known that this type of microfiber is rapidly destroyed in the study of metals and caustic substances,we investigated the lifetime of microfiber with gold nanoparticles deposited on it and using graphene films and nanoparticles. When applying gold nanoparticles with a diameter of 30 nm, the lifetime of the microfiber is 7 seconds. When graphene and gold nanoparticles were applied to microfiber, the microfiber lifetime was increased to 32 seconds. From the data obtained, we conclude that the presence of graphene films on the surface of microfiber increases its lifetime times 4.57. Further optimization of the concentration may lead to an even longer lifetime of the fiber.

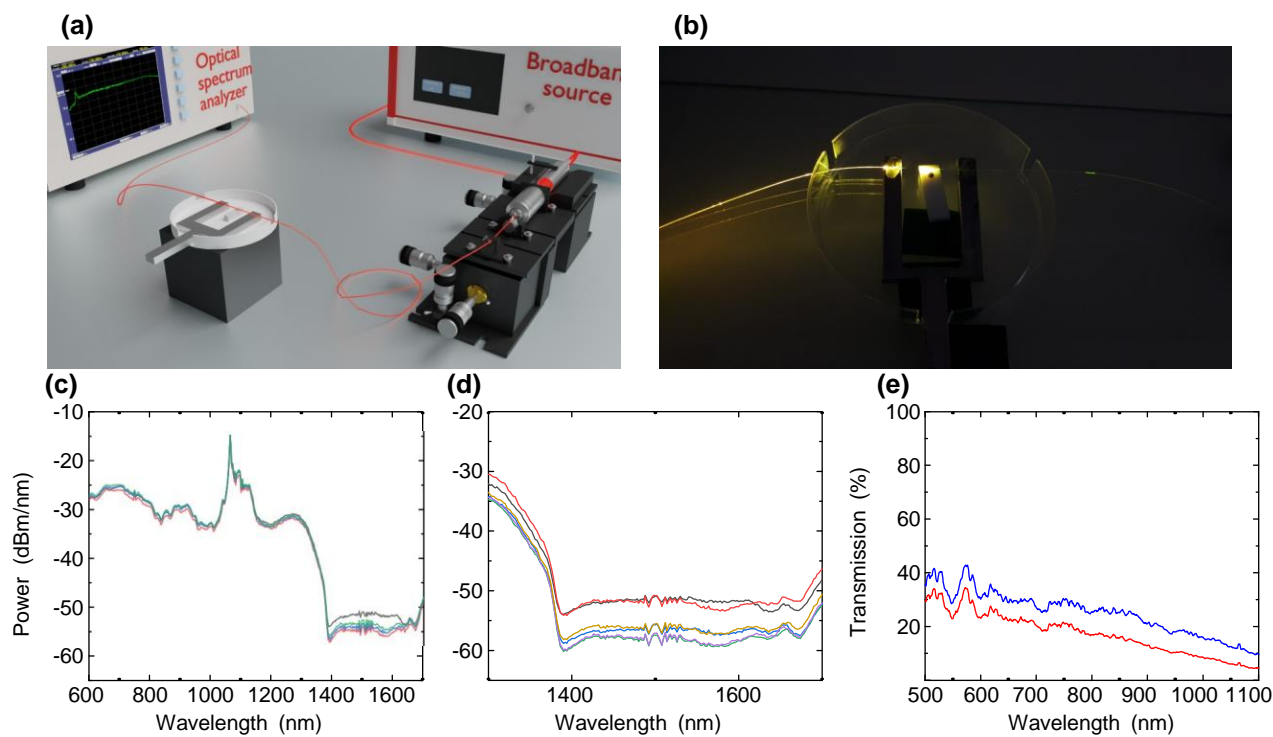

Figure 5: (a) Experimental setup used in the tapered fiber experiment. (b) The tapered fiber structure with the sample on the Teflon spacer. (c) VIS-NIR spectra of graphene and graphene with gold nanoparticles using cuvette at wavelength range of 600-1700 nm and (d) zoom-in intersection region at wavelength range of 1300-1700 nm. (e) Spectra of graphene overlayers on tapered fibers based on graphene form of shungite and gold nanoparticles at wavelength range of 500-1100 nm.

## 3   Conclusions

We examined the use of graphene films from the natural mineral shungite with the addition of gold nanoparticles of gold to increase the lifetime of microfiber (SMF-28). From the data obtained, we can conclude that the presence of graphene particles on the surface of microfiber increases its lifetime by 4.57 times. This method opens up the prospect of utilising microfibers with graphene for sensing and for the studies of corrosive substances and metals.

We conducted a study of graphene composites and selected the optimal ratio of gold nanoparticles to graphene (gold concentration 600 $\mu l$ - 1:3, graphene concentration 200 $\mu l$) at which the peaks of gold are clearly visible, but the microfiber does not degrade quickly and this gives time for a more detailed study of the peaks before the destruction of microfiber takes place.

This type of microfiber with composites applied to it can be used for research in various industries, such as environmental protection (determination of various pollutants in water, soil and air) , biotechnology and biology (determination of nucleotide sequences, proteins, viruses, bacteria), chemical industry (determination of new substances with low concentrations), pharmacology (study of new drugs with low concentrations), chemical industry (study new materials), construction industry (measurements of deformation and stress of materials, detection and characterization of damage, a new tool for measuring temperature and pressure), stratigraphy (probe for rapid snow grain size, determination of layered, sedimentary and volcanogenic rocks), manufacturing of smart materials, it can also be used to capture sound waves.

Wiley-VCH## 4 Experimental Section

*Fabrication of the sensing device*:
In our work, we used single-mode fiber (SMF-28) and tapered it using a commercial Vytran GPX-3000 tapering system. In the first step, we removed the acrylate coating from the fiber section where the tapering was carried out. After the removal of the coating, the fiber narrowed from a diameter of 125 $\mu m$ to a diameter of $\sim$15 $\mu m$. In the next step, the fiber was narrowed to a diameter of $\sim$2.5 $\mu m$. The fiber was then attached to a metal fork with epoxy glue to prevent the fiber from breaking during movement. To apply and hold the graphene samples with gold nanoparticles, we placed a Teflon spacer under the fiber.

*Obtaining composite materials*:
0.01 ml of nanoparticles of gold with a diameter of 30 nm was added to plastic eppendorfs with a capacity of 5 ml, then 4.5 ml of distilled water was added. Dispersion of nanoparticles was carried out using a Sony ultrasonic bath in a solution of distilled water. Plastic tubes with the obtained suspensions were placed in an ultrasonic bath. Having dispersed the nanoparticles, we broke up the microagglomerates, increasing the specific surface area of the nanoparticles. In flasks with a capacity of 50 ml, 25 ml of distilled water and 4.51 ml of dispersed gold nanoparticles were added. Then, 1 g of shungite powder was placed in each flask and placed in an ultrasonic bath for 1 hour. The resulting solutions were stored at room temperature.

*Application of a composite material to the microfiber surface*:
We applied 6 micro liters of composite material (graphene films + gold nanoparticles) to the obtained cone-shaped fiber with an automatic pipette, then we connected the laser. Under the thermal action of the laser, the water from the composite evaporated and the particles settled on the surface of the microfiber.

**Acknowledgements**
We are thankful to Israel Science Foundation (ISF) grant No. 2598/20 for supporting our research.

## References

[1] K. Kao, G. A. Hockham, In *Proceedings of the Institution of Electrical Engineers*, volume 113. IET, **1966** 1151–1158.

[2] S. Miller, *Optical fiber telecommunications*, Elsevier, **2012**.

[3] A. Karabchevsky, A. Katiyi, A. S. Ang, A. Hazan, *Nanophotonics* **2020**, *9*, 12 3733.

[4] L. Tong, F. Zi, X. Guo, J. Lou, *Optics Communications* **2012**, *285*, 23 4641.

[5] S. Wang, J. Wang, G. Li, L. Tong, *Applied optics* **2012**, *51*, 15 3017.

[6] D. A. Krohn, T. MacDougall, A. Mendez, *Fiber optic sensors: fundamentals and applications*, Spie Press Bellingham, WA, **2014**.

[7] E. Udd, W. B. Spillman Jr, *Fiber optic sensors: an introduction for engineers and scientists*, John Wiley & Sons, **2011**.

[8] E. Udd, *review of scientific instruments* **1995**, *66*, 8 4015.

[9] W. B. Ji, S. H. K. Yap, N. Panwar, L. L. Zhang, B. Lin, K. T. Yong, S. C. Tjin, W. J. Ng, M. B. A. Majid, *Sensors and Actuators B: Chemical* **2016**, *237* 142.

[10] J.-h. Li, J.-h. Chen, F. Xu, *Advanced Materials Technologies* **2018**, *3*, 12 1800296.

[11] A. Karabchevsky, A. Katiyi, M. I. M. Bin Abdul Khudus, A. V. Kavokin, *ACS Photonics* **2018**, *5*, 6 2200.

[12] X. Wu, L. Tong, *Nanophotonics* **2013**, *2*, 5-6 407.

[13] R. Black, E. Gonthier, S. Lacroix, J. Lapierre, J. Bures, In *Components for Fiber Optic Applications II*, volume 839. International Society for Optics and Photonics, **1988** 2–19.

[14] A. Katiyi, A. Karabchevsky, *Journal of Lightwave Technology* **2017**, *35*, 14 2902.

[15] J. I. Peterson, G. G. Vurek, *Science* **1984**, *224*, 4645 123.

[16] M. Gandhi, S. Chu, K. Senthilnathan, P. R. Babu, K. Nakkeeran, Q. Li, *Applied Sciences* **2019**, *9*, 5 949.

[17] N. A. S. Omar, Y. W. Fen, J. Abdullah, Y. M. Kamil, W. M. E. M. M. Daniyal, A. R. Sadrolhosseini, M. A. Mahdi, *Scientific reports* **2020**, *10*, 1 1.

[18] H. Jia, A. Zhang, Y. Yang, Y. Cui, J. Xu, H. Jiang, S. Tao, D. Zhang, H. Zeng, Z. Hou, et al., *Lab on a Chip* **2021**.

[19] X. Li, H. Ding, *IEEE Photonics Technology Letters* **2014**, *26*, 16 1625.

[20] Y. Li, H. Ma, L. Gan, A. Gong, H. Zhang, D. Liu, Q. Sun, *Journal of biophotonics* **2018**, *11*, 9 e201800012.

[21] X. Liu, W. Lin, P. Xiao, M. Yang, L.-P. Sun, Y. Zhang, W. Xue, B.-O. Guan, *Chemical Engineering Journal* **2020**, *387* 124074.

[22] Y. Huang, P. Chen, H. Liang, A. Xiao, S. Zeng, B.-O. Guan, *Biosensors and Bioelectronics* **2020**, *156* 112147.

[23] A. Katiyi, J. Zorea, A. Halstuch, M. Elkabets, A. Karabchevsky, *Biosensors and Bioelectronics* **2020**, *161* 112240.

[24] W. Zhou, K. Li, Y. Wei, P. Hao, M. Chi, Y. Liu, Y. Wu, *Biosensors and Bioelectronics* **2018**, *106* 99.

[25] M. El-Sherif, L. Bansal, J. Yuan, *Sensors* **2007**, *7*, 12 3100.

[26] D. Monzón-Hernández, D. Luna-Moreno, D. Martínez-Escobar, J. Villatoro, In *2nd Workshop on Specialty Optical Fibers and Their Applications (WSOF-2)*, volume 7839. International Society for Optics and Photonics, **2010** 78390I.

[27] H. Fu, Y. Jiang, J. Ding, J. Zhang, M. Zhang, Y. Zhu, H. Li, *Sensors and Actuators B: Chemical* **2018**, *254* 239.

[28] V. Kurevin, O. Morozov, V. Prosvirin, A. Salihov, A. Smirnov, *Infokommunikacionnye tehnologii* **2009**, *7*, 3 46.

[29] J. S. Selker, Taking the temperature of ecological systems with fiber optics: Fiber optic distributed temperature sensing for ecological characterization; blue river, oregon, 10–15 september 2007, **2008**.

[30] S. Kholodkevich, V. Fedotov, A. Ivanov, T. Kuznetsova, A. Kurakin, E. Kornienko, *terrorism* **2007**, *1* 2.

[31] M. Dai, Z. Chen, Y. Zhao, M. S. Aruna Gandhi, Q. Li, H. Fu, *Biosensors* **2020**, *10*, 11 179.

[32] A. D. Kersey, A. Dandridge, *IEEE Transactions on components, hybrids, and manufacturing technology* **1990**, *13*, 1 137.

[33] A. Nanni, C. Yang, K. Pan, J. S. Wang, R. R. Michael, *ACI materials Journal* **1991**, *88*, 3 257.

[34] J. Nedoma, M. Stolarik, M. Fajkus, M. Pinka, S. Hejduk, *Applied Sciences* **2019**, *9*, 1 134.

[35] J. Gerlici, O. Nozhenko, G. Cherniak, M. Gorbunov, R. Domin, T. Lack, In *MATEC Web of Conferences*, volume 157. EDP Sciences, **2018** 03007.

[36] B. Glisic, D. L. Hubbell, D. H. Sigurdardottir, Y. Yao, *Optical Engineering* **2013**, *52*, 8 087101.

[37] S. Großwig, E. Hurtig, K. Kühn, *Geophysics* **1996**, *61*, 4 1065.

[38] S. W. Tyler, J. S. Selker, M. B. Hausner, C. E. Hatch, T. Torgersen, C. E. Thodal, S. G. Schladow, *Water Resources Research* **2009**, *45*, 4.

*[39]* M. A. El-Sherif, J. Yuan, A. MacDiarmid, *Journal of intelligent material systems and structures* **2000**, *11*, 5 407.

*[40]* D. F. Berisford, N. P. Molotch, M. T. Durand, T. H. Painter, *Cold regions science and technology* **2013**, *85* 183.

[41] S. W. Tyler, S. A. Burak, J. P. McNamara, A. Lamontagne, J. S. Selker, J. Dozier, *Journal of Glaciology* **2008**, *54*, 187 673.

[42] S.-M. Chuo, L. A. Wang, *Optics Communications* **2011**, *284*, 12 2825.

[43] A. Novikova, A. Karabchevsky, *arXiv preprint arXiv:2110.12790* **2021**.

**Table of Contents**

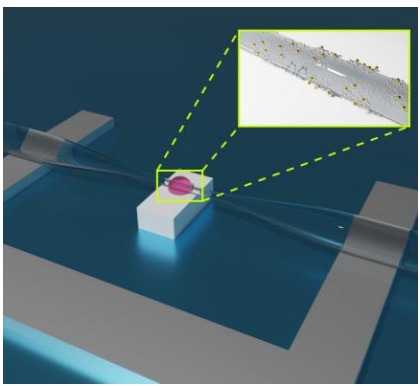

ToC Entry - Optical microfibers are finding new applications in various fields of industry, which, in turn, requires high resistance, environmental friendliness and ease of use. Here we report a new method to prolong the microfiber lifetime by modifying its surface with green-extracted graphene overlayers which increase the lifetime and survivability of the microfiber 5 times, as compared to the bare microfiber.